\documentclass[sigconf, pbalance=true]{acmart}

\usepackage{longtable}
\usepackage{booktabs}
\usepackage{graphicx}
\usepackage{amsmath}
\usepackage{placeins}
\usepackage{needspace} 
\usepackage{subcaption}

\AtBeginDocument{%
  }

\copyrightyear{2026}
\acmYear{2026}
\setcopyright{cc}
\setcctype{by}
\acmConference[CHI EA '26]{Extended Abstracts of the 2026 CHI Conference on Human Factors in Computing Systems}{April 13--17, 2026}{Barcelona, Spain}
\acmBooktitle{Extended Abstracts of the 2026 CHI Conference on Human Factors in Computing Systems (CHI EA '26), April 13--17, 2026, Barcelona, Spain}
\acmDOI{}
\acmISBN{}

\begin{document}

\title{Feeds Don't Tell the Whole Story: Measuring Online-Offline Emotion Alignment}

\author{Sina Elahimanesh}
\authornote{Both authors contributed equally to this research.}
\affiliation{%
  \institution{Saarland University}
  \city{Saarbrücken}
  \country{Germany}
}
\email{siel00002@uni-saarland.de}

\author{Mohammadali Mohammadkhani}
\authornotemark[1]
\authornote{This author is supported by the Konrad Zuse School of Excellence in Learning and Intelligent Systems (\textcolor{blue}{\href{https://eliza.school/}{ELIZA}}) through the DAAD programme Konrad Zuse Schools of Excellence in Artificial Intelligence, sponsored by the Federal Ministry of Education and Research.}
\affiliation{%
  \institution{Saarland University}
  \city{Saarbrücken}
  \country{Germany}
}
\affiliation{
    \institution{Zuse School ELIZA}
    \country{Germany}
}
\email{momo00016@stud.uni-saarland.de}

\author{Shohreh Kasaei}
\affiliation{
  \institution{Sharif University of Technology}
  \city{Tehran}
  \country{Iran}
}
\email{kasaei@sharif.edu}

\renewcommand{\shortauthors}{Elahimanesh et al.}

\begin{abstract}
  A clear and well-documented \LaTeX\ document is presented as an
  article formatted for publication by ACM in a conference proceedings
  or journal publication. Based on the ``acmart'' document class, this
  article presents and explains many of the common variations, as well
  as many of the formatting elements an author may use in the
  preparation of the documentation of their work.
\end{abstract}



























\begin{abstract}
In contemporary society, social media is deeply integrated into daily life, yet emotional expression often differs between real and online contexts. We studied the Persian community on X to explore this gap, designing a human-centered pipeline to measure alignment between real-world and social media emotions. Recent tweets and images of participants were collected and analyzed using Transformers-based text and image sentiment modules. Friends of participants provided insights into their real-world emotions, which were compared with online expressions using a distance criterion. The study involved N=105 participants, 393 friends, over 8,300 tweets, and 2,000 media images. Results showed only 28\% similarity between images and real-world emotions, while tweets aligned about 76\% with participants' real-life feelings. Statistical analyses confirmed significant disparities in sentiment proportions across images, tweets, and friends' perceptions, highlighting differences in emotional expression between online and offline environments and demonstrating practical utility of the proposed pipeline for understanding digital self-presentation.

\end{abstract}

 \maketitle

\renewcommand{\thefootnote}{\textasteriskcentered}

\section{Introduction}
\label{sec:intro}

\begin{figure*}
\centering
\includegraphics[width=\textwidth]{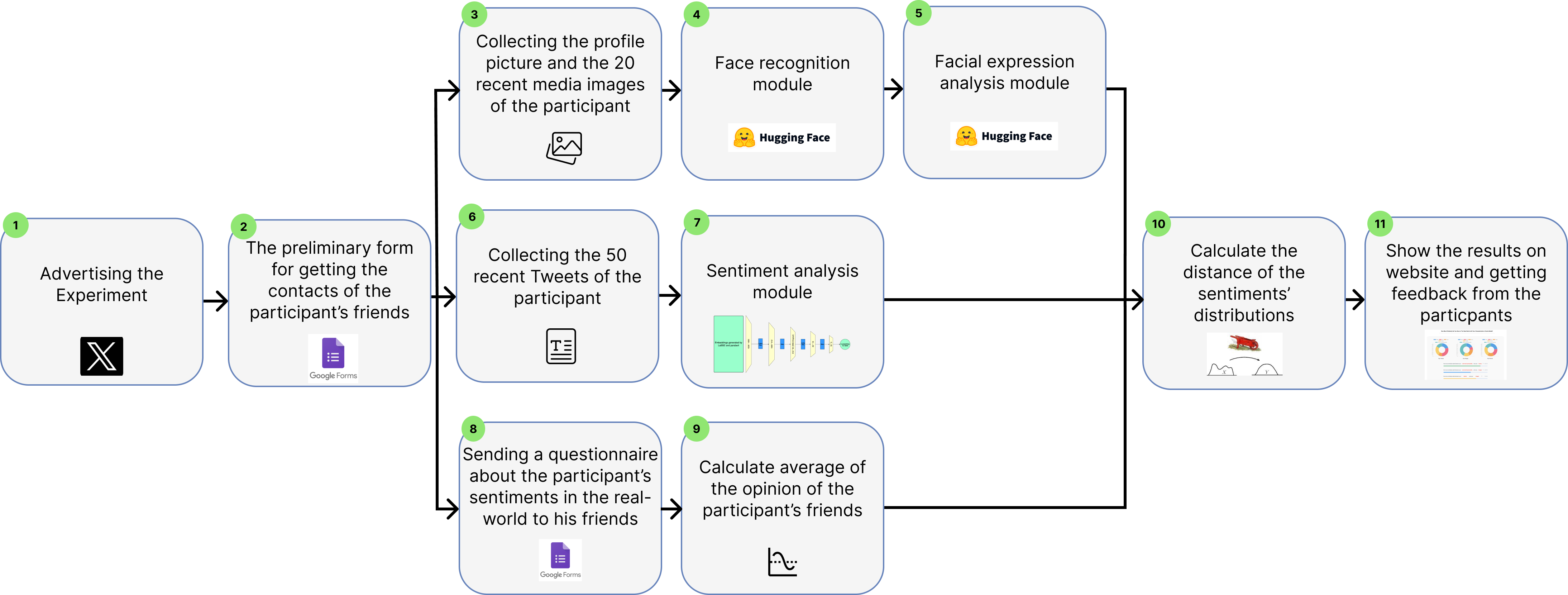}
\caption{Overview of the proposed experimental pipeline. The workflow includes participant recruitment, collection of tweets and images, AI-based sentiment and facial expression analysis, aggregation of friends’ offline assessments, computation of sentiment distribution distances, and visualization of alignment results.}
\label{fig:pipeline}
\end{figure*}

Social networking platforms have become central to daily life, particularly among youth who frequently use them \cite{twenge2019trends}. A key motivation for engagement is expressing viewpoints and emotions through textual or multimedia content \cite{bala2014social}. Technological advances, particularly smartphones and widespread Internet access, have shifted everyday communication onto social networks \cite{srivastava2005mobile}, increasing reliance on these platforms. In response, social networks have developed applications and messaging tools to facilitate communication \cite{ortiz2023rise, appel2020future}, encouraging engagement \cite{claussen2013effects} and extending users' online presence \cite{scott2017time}. Consequently, social media has become a rich source for examining people's behaviors, social interactions, and virtual characteristics.

However, users may not always convey their true emotions and personalities online \cite{balfe2010disclosure}, and they may express different emotions to different audiences \cite{zheng2020self, vitak2012impact}. Factors such as shyness or apprehension about revealing genuine feelings contribute to these discrepancies \cite{fang2017coping}. Thus, social media cannot serve as the sole lens for understanding personality but can provide complementary insights when combined with real-world data.

Emotions are a fundamental aspect of personality \cite{montag2017primary}; analyzing them across contexts offers a nuanced understanding of human behavior. Emotional expression varies with environment and context \cite{hess2020bidirectional}, and social norms further shape how individuals express feelings \cite{fischer2003social}. People may exhibit distinct emotional patterns depending on the situation \cite{sherman2015independent}. By comparing emotions across online and offline environments, this study addresses RQ \ref{RQ}. While some research investigates links between real-world behaviors and social media emotions \cite{mirlohi2021causal}, evidence on alignment or divergence is limited. We address this gap through a hybrid pipeline combining human insights and AI-based sentiment analysis, aiming to illuminate emotional expression in digital spaces and guide HCI research in understanding online-offline emotional dynamics. We propose the following research question (RQ \ref{RQ}): \textbf{RQ: Do people express the same emotions on social media as they experience in real life?}\label{RQ}


To investigate this question, we designed a human-centered pipeline combining AI-based analysis with real-world assessments. Participants' tweets and posted images were analyzed using text and image sentiment modules to estimate online emotions, while friends familiar with the participants provided offline emotional assessments. Statistical tests and distribution-distance metrics were used to quantify differences across modalities. Results were visualized through an interactive website designed for transparency while addressing ethical and privacy considerations. We focused on Persian-speaking users on \textit{X}, a large and active community \cite{zhou2010information}. To define emotion categories, we conducted interviews with ten active users, resulting in five labels: \textit{Happiness}, \textit{Sadness}, \textit{Anger}, \textit{Neutral}, and \textit{Intense Emotions} (balancing positive, negative, and neutral expressions).

Our contributions are threefold: \textbf{(1)} we created a novel Persian X dataset of over 3000 labeled tweets and images (published in Huggingface \footnote{https://huggingface.co/datasets/moali-mkh-2000/PersianTwitterDataset-SentimentAnalysis}), with a Transformers-based text sentiment model achieving over 74\% accuracy; \textbf{(2)} we developed a human-centered pipeline integrating AI-based modules with real-world assessments to measure online-offline emotional alignment; \textbf{(3)} we conducted an experiment with N=105 participants and 393 friends, showing ~76\% similarity between tweets and real-life emotions, but only ~29\% similarity for images. To address potential subjectivity in friends' assessments, we collected self-reports at the end of the study, and about 93\% of participants agreed that the analysis accurately reflected both their offline and online emotions.

\section{Related Work}

Our work builds on research examining how online emotional expression relates to individuals' real-world affective states. Social computing studies show that digital traces can capture aspects of users' offline personality, behavior, and well-being \cite{staiano2012friends, selfhout2010emerging, keles2020systematic}, motivating efforts to infer psychological traits from online interactions \cite{qiu2012you, celli2011mining, tai2016systematical}. Yet HCI scholarship consistently highlights the mismatch between felt and expressed emotion in networked environments. A CHI-published study \cite{saha2021life} demonstrates that people disclose only selective and predominantly positive life events on social media, while Andalibi and Forte \cite{andalibi2018responding} show that emotional expression is shaped by audience expectations and sociotechnical norms. Foundational work on online identity further illustrates how individuals strategically manage self-presentation \cite{zhao2008identity}, and workplace studies highlight how concerns about visibility and context constrain emotional disclosure \cite{vitak2014you}. Collectively, these findings indicate that social media provides only a partial view of people's emotional lives, underscoring the need for approaches that relate online signals to offline assessments.

Advances in computational methods have further transformed the analysis of emotional content. Early sentiment analysis relied on feature-based machine-learning models such as SVMs and Naive Bayes \cite{golbeck2011predicting, celli2012adaptive}, whereas recent deep learning and transformer-based approaches offer substantial gains in text and image emotion recognition \cite{abbas2021social, vaswani2017attention, zhang2023sentiment}. These techniques now underpin a broad range of applications in social media analytics, including behavioral modeling, event monitoring, and content moderation \cite{cui2018travel, aljarah2021intelligent}.

On X (Twitter), research has evolved from traditional classifiers \cite{hasan2018machine, pratama2015personality} to transformer-based models addressing tasks such as hate-speech detection, pandemic-related discourse, and sarcasm recognition \cite{naseem2020transformer, shi2023bert, sarsam2020sarcasm}. Work on Persian-language content has similarly advanced, with deep learning approaches for sentiment analysis \cite{fatehimbti, heydari2021deep} and specialized tasks including emoji prediction, rumor detection, and political behavior mapping \cite{tavan2020persian, sadr2020predictive, khazraee2019mapping}. Extending this literature, our study introduces a human-centered multimodal framework that directly compares Persian users' online emotional expressions with offline assessments provided by close social ties, a perspective largely absent from prior research.

\section{Proposed Method}

\begin{figure*}[h!]
\centering
  \includegraphics[width=\textwidth]{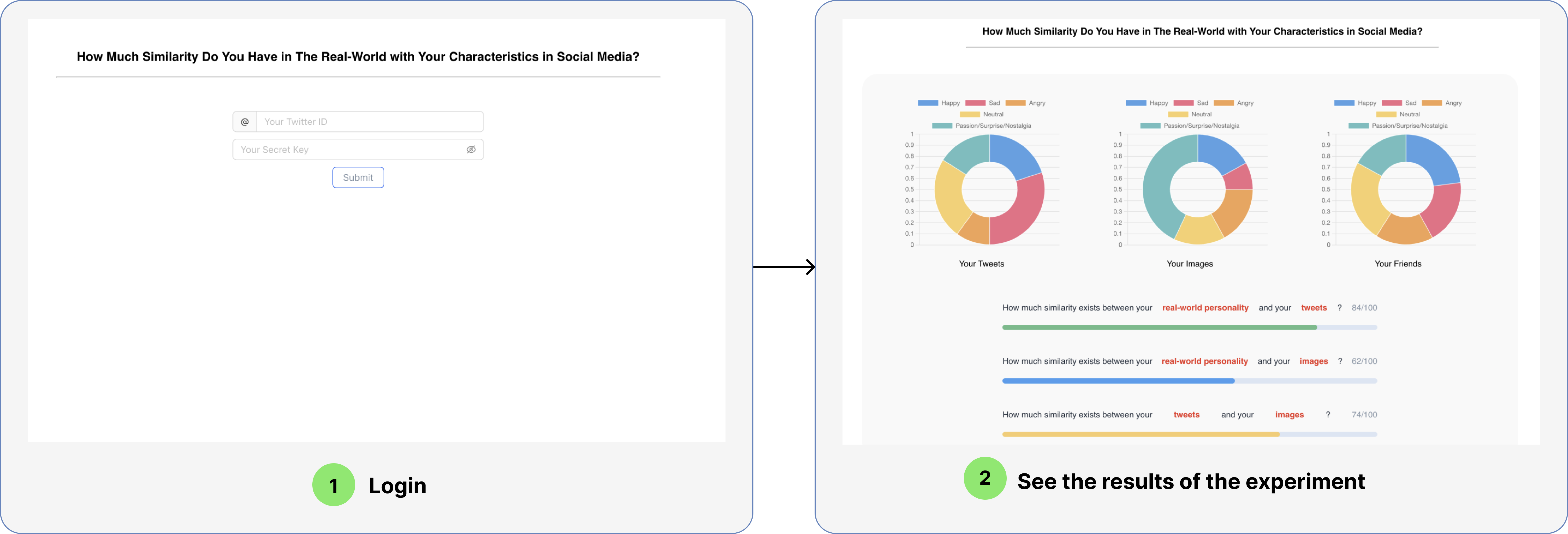}

  \caption{Overview of the experimental website used in the post-survey study: participants first log in with their X platform ID and a secret key to access their social media data (Step 1), and then view the analysis results (Step 2), where donut charts show the distribution of emotional tones (e.g., happy, sad, neutral, angry) and bar charts present similarity scores (0–100) comparing real-world personality with expressions across tweets, uploaded images, and friends’ content.}

  \label{app:website}

\end{figure*}


\textit{\textbf{An Overview of The Proposed Method.}}
Our pipeline consists of \textit{11 stages} (Figure \ref{fig:pipeline}). In \textit{Stage 1}, the experiment is advertised on X via an invitation tweet, generating the main participant pool. \textit{Stage 2} collects participants' basic information and friends' contacts through a Google Form, with detailed explanations and consent obtained for crawling tweets and images.
In \textit{Stages 3 and 6}, 50 recent tweets and the profile image plus 20 recent images are collected; participants must have at least this amount of content. \textit{Stage 8} gathers friends' quantitative assessments of participants' real-world emotions, and \textit{Stage 9} averages these as representative sentiment distributions.
Images pass through a face recognition module (\textit{Stage 4}) and a facial expression module (\textit{Stage 5}) to classify sentiments, while tweets are processed by a sentiment analysis module (\textit{Stage 7}). \textit{Stage 10} computes distances between real-world, tweets, and images sentiment vectors. Finally, in \textit{Stage 11}, participants view results on a website and provide qualitative and quantitative feedback.

\textit{\textbf{Real-World Sentiments Extraction.}}
This stage extracts real-world sentiments by surveying individuals close to the participant rather than the participant themselves, avoiding self-bias \cite{van2008faking, ong2018happier}. Participants select 3-5 close contacts, who receive a form with metadata and five questions, each targeting one of the five sentiment categories. Questions focus on recent behaviors, with responses rated on a 1-10 scale, ensuring alignment with recent tweets and images. The collected data generate a sentiment distribution normalized into probabilities, leveraging friends' understanding of the participant's emotional tendencies. An example question is: \textbf{How frequently or to what extent do you believe your friend has been Happy in the real world recently? (1-10)}.

\textit{\textbf{Dataset Gathering and Preprocessing.}}
As no suitable Persian X dataset existed, we created one by collecting over 3,300 tweets from Persian-speaking users using the Snscrape\footnote{https://github.com/JustAnotherArchivist/snscrape}
 package. The first two authors independently annotated all records into five labels; only 0.1\% required resolution through discussion. Texts were normalized using the \textit{Hazm} Python package to standardize tweets, and Pandas was used for EDA to detect duplicates or empty entries. After anonymization, normalization, and labeling, the dataset was ready for text analysis models.
. 



\textit{\textbf{Text Sentiment Analysis Pipeline.}}
Embeddings are generated using two models. ParsBERT \cite{ParsBERT}, a pre-trained Persian BERT, produces a 768-dimensional vector per input, and LaBSE \cite{feng2020language}, a multilingual model, produces another 768-dimensional vector. These embeddings are concatenated into a 1536-dimensional vector per tweet. A fully connected neural network maps each vector to five probabilities corresponding to the labels, selected over alternatives such as SVM, Naïve Bayes, and Random Forest due to better performance.
To improve performance, we defined keyword lists per label. These keywords frequently appear in specific sentiments. After the classifier predicts a label and PyTorch provides a confidence score, if confidence is at least 80\%, the predicted label is used. Otherwise, the system checks for keywords; if a relevant keyword is found, its label is assigned; if not, the classifier's label is used. The model achieves \textbf{accuracy=74.08}\%, with \textbf{F1=72.02}\%, \textbf{precision=72.03}\%, and \textbf{recall=75.99}\%.

\textit{\textbf{Image Sentiment Analyses Pipeline.}}
The image pipeline includes face recognition and facial expression analysis. A Hugging Face face recognition model\footnote{https://huggingface.co/spaces/ParisNeo/FaceRecognition}
 uses the participant's profile image to locate their face in collected images. A second Hugging Face model\footnote{https://huggingface.co/spaces/schibsted/Facial\_Recognition\_with\_Sentiment\_Detector}
, based on VGG19 trained on FER2013, detects sentiments in these facial images. Its seven basic emotions are mapped to five categories by merging \textit{Sad} with \textit{Fear} and \textit{Angry} with \textit{Disgust}. The final output is an averaged probability vector of sentiments across all images.

\begin{table*}[t]
\centering
\caption{Experiment results, showing mean and STD of each sentiment. Real-world and tweet emotions are similarly distributed, while profile/media images are predominantly classified as happiness or intense emotions.}
\begin{tabular*}{\textwidth}{@{\extracolsep{\fill}}lccc@{}}
\toprule
\textbf{Sentiment} & \textbf{Real World} & \textbf{Tweets' Text} & \textbf{Profile and Media Images} \\ 
\midrule
\textbf{Happiness}        & 24.37\% (± 6.79\%)  & 16.79\% (± 8.75\%) & 48.40\% (± 21.65\%) \\
\textbf{Sadness}          & 18.11\% (± 6.14\%)  & 22.65\% (± 9.96\%) & 8.36\%  (± 8.44\%) \\
\textbf{Neutral}          & 20.14\% (± 6.02\%)  & 24.36\% (± 9.87\%) & 3.94\%  (± 5.68\%) \\
\textbf{Anger}            & 13.67\% (± 4.84\%)  & 15.76\% (± 7.85\%) & 4.50\%  (± 4.42\%) \\
\textbf{Intense Emotions} & 22.78\% (± 5.35\%)  & 20.46\% (± 8.27\%) & 34.84\% (± 18.57\%) \\
\bottomrule
\end{tabular*}
\label{tab:sentiments}
\end{table*}

\textbf{\textit{Sentiment Comparison Metric.}}
The previous steps produce three discrete distributions over five sentiment classes (tweets, images, and real-world reports). To compare them, we compute pairwise distances. The distance metric must detect both subtle and large differences while remaining bounded, allowing similarity scores to be mapped to a 0--100 scale for participant-friendly reporting. We considered Jensen--Shannon Divergence and KL-Divergence, but selected the \textit{Earth Mover's Distance} (EMD) because it satisfies these requirements. For discrete ordered classes, we use
$
D_{\text{EMD}}(P, Q) = \sum_i |P(i) - Q(i)|,
$
where \(P\) and \(Q\) denote the sentiment distributions. EMD~\cite{andoni2008earth} offers sensitivity to small shifts, bounded outputs, and captures the minimal ``effort'' required to transform one distribution into the other.

\textbf{\textit{Feedback Collection Website.}}
We designed a website with a Django backend and a React frontend. Django handles backend logic; React manages the user interface. Responsiveness was required so the design remains consistent across devices. After development, both backend and frontend were deployed on a server, and the website address was sent to participants.
The website flow is as follows: participants enter their ID on X and credentials. If correct, they are directed to the main screen where results are shown in three charts representing tweets, images, and real-world opinions. Below the charts, three similarity scores (0-100) report pairwise similarities between tweets, images, and friends' assessments, indicating alignment between emotions on X and in real life.

\textbf{\textit{Ethics and Privacy Concerns.}}
Participants were fully informed about the experiment's purpose and procedures, including collection and publication of anonymized tweet data. Explicit consent covered use of anonymized sentiment distributions, distance values, and friends' data for supplementary analysis; no incentives were provided. Privacy was ensured by assigning users passwords to view results, anonymizing tweets, and safeguarding identities. Participants did not know which friend provided real-life emotion assessments, protecting friends' privacy. At the end, participants completed a feedback form; all comments were analyzed, no ethical issues arose, and participants were satisfied with the procedure and handling of ethical and privacy concerns.

\section{Experimental Evaluation and Results}

The experiment involved N=105 Persian-speaking participants aged 17-33 (mean 21.84, SD=2.13; 63 males, 42 females). For each, 50 recent tweets and 20 profile/media images were collected. Tweets were analyzed with the text sentiment module to compute sentiment distributions. A Google Form was sent to 3-5 friends per participant (avg. 3.74, total 393, all over 16). Table \ref{tab:sentiments} shows average sentiment distributions for real-world emotions, social media images, and tweets. Participants appeared happiest or most passionate in photos, while tweets showed less happiness. Real-world emotions were dominated by happiness, anger least common; tweets were mostly neutral; images were mostly happy.

\textbf{\textit{Sentiment Similarities.}}
Average similarity between tweet sentiments and real-world emotions was ~76\%, showing tweets reliably reflect real feelings. Image similarity was lower (~29\%), as people often appear happier in photos. Similarity between image and tweet sentiments was ~38\%, highlighting the disconnect between these mediums (Table~\ref{tab:sentiments_similarity}).





\begin{table}[t]
\centering
\caption{Mean similarity (\%) between sentiment distributions across modalities (complement of Earth Mover’s Distance).}
\begin{tabular*}{\columnwidth}{@{\extracolsep{\fill}}lcc}
\toprule
\textbf{Comparison} & \textbf{Mean} & \textbf{SD} \\
\midrule
Tweets vs.\ Real-world & 75.88\% & 14.69\% \\
Images vs.\ Real-world & 28.67\% & 30.01\% \\
Images vs.\ Tweets     & 38.02\% & 29.78\% \\
\bottomrule
\end{tabular*}
\label{tab:sentiments_similarity}
\end{table}

\textit{\textbf{Statistical Analysis on Sentiment Similarities.}}
To validate differences in sentiment distributions (images, tweets, friends' perceptions), we conducted independent two-sample t-tests. Pairwise comparisons were made for each sentiment between images, tweets, and friends' perceptions, with the null hypothesis assuming no difference in mean proportions. Results report t-statistics, p-values, and significance at $\alpha=0.05$. All comparisons were statistically significant with very low p-values, confirming that emotional expressions differ across mediums and that external perceptions vary.

Although all pairwise comparisons were statistically significant at $\alpha=0.05$, we did not apply a formal multiple-comparison correction (e.g., Bonferroni or FDR). As multiple t-tests increase the risk of Type I error, results should be interpreted with caution; however, the extremely small p-values and consistent effect directions across sentiments suggest that the observed differences are unlikely to be driven solely by multiple testing.

\textit{\textbf{Post-Study's Qualitative and Quantitative Analysis.}}
Participants received website credentials and a Google Form with quantitative and qualitative questions. About 93\% rated the experiment as accurately analyzing their emotions. Qualitative feedback, categorized into 10 clusters, showed satisfaction with user experience, analysis accuracy, and questionnaire design, and appreciation for awareness of their social media emotions. Suggestions included more diverse tweets/images, time-based crawling, multi-platform extension, and personalization for introverts versus extroverts. Participants valued privacy, ethics, and transparency. Overall, the feedback provides insights on satisfaction, challenges, and improvements, validating the study and guiding future work.

\section{Discussion}

This research is motivated by prior work (Section \ref{sec:intro}) highlighting emotions as a core component of personality and noting that emotional expression varies across contexts. Given social media's central role in shaping interpersonal and digital experiences, we aimed to quantify divergence between real-life and social media emotions. Despite a lack of comparable studies, we developed and post-evaluated a framework grounded in human-centered methods. Similar works \cite{kuvsen2017identifying, gaind2019emotion} differ in that our pipeline actively involves users, focusing on personality and emotional experiences across real and digital contexts, whereas prior works primarily analyze social media emotions in isolation.

\textit{\textbf{Theoretical and Practical Contributions.}}
We contribute both theoretically and practically by examining how people express emotions across real and digital contexts. A novel dataset of Persian tweets was collected and labeled into 5 emotion categories. A hybrid model combining a Transformers-based Persian sentiment module with a rule-based component achieved about 75\% accuracy. Leveraging this module alongside facial expression analysis and human input, a pipeline was introduced to quantify alignment between offline and online emotional expressions. Statistical analyses confirmed significant differences in emotion proportions. A user study with N=105 participants and 393 acquaintances collected over 5000 tweets and 2000 images. Post-study evaluation using qualitative and quantitative methods indicated about 93\% satisfaction, with participants highlighting limitations that guide future research.

\textbf{\textit{Findings.}}
Our findings reveal substantial gaps between individuals' real-world emotions and their online representations. The pipeline effectively surfaces these gaps and appears adaptable to other platforms and languages, enabling broader explorations of emotional communication in digital contexts. The study addressing RQ \ref{RQ} shows that tweets reflect offline emotions with relatively high fidelity, whereas images often misrepresent feelings. Textual content thus provides a more reliable signal of emotional states than visual content, and the statistical significance of differences emphasizes the role of medium in emotion analysis. However, the low alignment between image-based sentiments and real-world emotions should not be interpreted as a failure of the image analysis pipeline. Rather, it likely reflects performative self-presentation practices on social media, where users strategically curate visually positive or socially desirable images, resulting in systematic divergence from offline emotional states. 

Furthermore, significant variations across images, tweets, and friends' perceptions indicate that each perspective contributes unique insights, supporting multi-modal approaches to achieve a more holistic understanding of digital emotional expression.
Beyond reporting magnitude differences, our results suggest that emotional alignment online is medium-dependent and socially situated. Platform affordances shape how authenticity is performed, with text enabling relatively direct affective disclosure and images functioning more strongly as curated identity signals. This contributes to HCI theories of digital self-presentation by demonstrating that emotional authenticity is not uniform across modalities but conditioned by socio-technical affordances.

\textbf{\textit{Social Media Design Implications.}}
Our study shows that online emotions often differ from real-life feelings. Social media designers could consider these gaps by creating interfaces that help users reflect on their emotional expression, for example through visual summaries of text and image activity. Using text, images, and self-reports together can help users understand their online self-presentation and support more mindful, authentic interactions.

\textbf{\textit{Limitations and Future Work.}}
Several limitations affect the generalizability and scope of our findings. First of all, while friends’ reports provide an external and socially grounded perspective on participants’ offline emotions, they do not constitute objective ground-truth. Such assessments may vary across raters and be shaped by relationship closeness, subjective interpretation, and cultural norms around emotional expression, which should be considered when interpreting the reported alignment scores. Future work could strengthen validation by collecting time-matched self-reports alongside peer assessments and AI-based analyses of tweets and images. Triangulating these three sources, including friends’ evaluations, self-reports, and computational sentiment measures, would allow estimation of inter-rater agreement and increase confidence in offline emotion measurement.

Furthermore, focusing on a single language and region limits applicability, as emotional expression is culturally situated and varies across contexts. Because our sample consists of Persian-speaking users, norms surrounding public visibility and emotional display on X may have shaped how participants curated textual and visual content, potentially contributing to the observed divergence across modalities. Future work should extend the framework to multiple languages and regions using analogous sentiment models to examine cross-cultural patterns and assess generalizability. Our study examined text and images, excluding other media such as GIFs, memes, and short videos that may convey emotions differently; incorporating these could provide richer perspective. Additionally, the image analysis focuses primarily on facial expressions, overlooking contextual cues, body language, and compositional elements that may convey emotion. This limited representation may partially contribute to the lower image–offline alignment observed.

Limiting analysis to a single platform, X, constrains generalizability to other social media with distinct norms and affordances; future research should consider multiple platforms. Predefined sentiment labels and specific models also may restrict outcome diversity, suggesting exploration of alternative labeling schemes and model architectures. Age distribution was narrow, and broader inclusion could capture age-related differences. While our approach employs robust models, integrating state-of-the-art architectures and expanded evaluations may improve accuracy. Finally, temporal dynamics and individual differences in emotional expression were not modeled; incorporating time-sensitive content and personalized pipelines could yield more nuanced insights.

\section{Conclusion}
This study, addressing potential discrepancies between real and virtual emotions, introduced a framework measuring alignment between social media and real-life emotions. The pipeline combines human perspectives with AI and was evaluated with N=105 participants. Post-study analysis showed 93\% satisfaction, validating the method. Statistical tests confirmed significant differences across images, tweets, and friends' perceptions, while qualitative analysis offered insights for future research.



\bibliographystyle{ACM-Reference-Format}

\bibliography{references}
















\clearpage

\appendix
\setcounter{page}{1}

\section*{Appendices}

\section{Comparison of Distribution Distance Metrics}

\label{appendix:distance_metrics}

\subsection{KL-Divergence}

KL-Divergence (Kullback-Leibler Divergence) \cite{hershey2007approximating} measures the distance between two probability distributions by summing the divergence of each distribution from the other. It is commonly used but does not handle small changes effectively and lacks boundedness, making it less suitable for our application. The formula used for calculation is:

\[
D_{KL}(P \parallel Q) = \sum_{i} P(i) \log \left( \frac{P(i)}{Q(i)} \right)
\tag{2}
\]

KL-Divergence can be problematic because it tends to give unbounded values, meaning that a small difference between two similar distributions can result in a large distance, making it hard to interpret. For example, comparing distributions \( P = [0.4, 0.1, 0.1, 0.2, 0.2] \) and \( Q = [0.3, 0.2, 0.2, 0.2, 0.1] \) yields a result of approximately 0.57, which fails to provide a meaningful measure of similarity for distributions that are closely related.

\subsection{Jensen-Shannon Divergence}

Jensen-Shannon Divergence \cite{menendez1997jensen} is a symmetrized and smoothed version of KL-Divergence. It improves upon KL-Divergence by being bounded, but it still does not capture small variations as effectively as Earth Mover's Distance. The formula used for calculation is:

\[
D_{JS}(P \parallel Q) = \frac{1}{2} \left[ D_{KL}(P \parallel M) + D_{KL}(Q \parallel M) \right]
\tag{3}
\]

where \( M = \frac{1}{2}(P + Q) \). While Jensen-Shannon Divergence ensures that the distance is bounded and symmetrized, it does not sufficiently capture small differences between distributions. In our example of comparing \( P = [0.4, 0.1, 0.1, 0.2, 0.2] \) and \( Q = [0.3, 0.2, 0.2, 0.2, 0.1] \), the result was about 0.06, which is not sufficiently sensitive to minor differences between the distributions. This makes it less effective for fine-grained comparisons in our study.

These two metrics, while useful in other contexts, did not meet the criteria for our analysis, as they either lacked boundedness or failed to capture small variations effectively. In contrast, using the same example, \textbf{Earth Mover's Distance} yielded a result of 0.4. This value reflects a bounded and sensitive measure of dissimilarity, making it the most appropriate metric for assessing the similarity between sentiment distributions.

\onecolumn

\section{Sample Tweets}

\begin{figure}[h]
\centering
  \includegraphics[width=\textwidth]{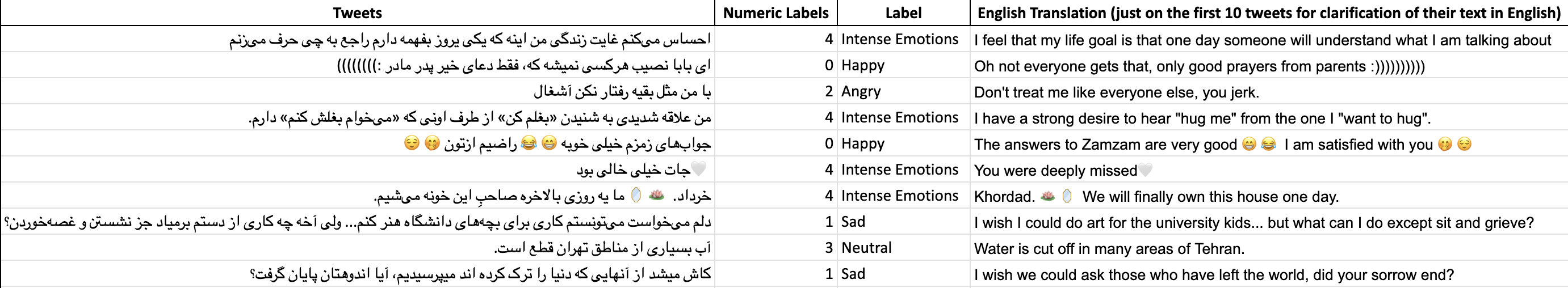}

  \caption{A screenshot of 10 tweets from the gathered dataset is shown here.}

  \label{app:sampletweets}

\end{figure}




\section{Sentiment Model Architecture and Evaluation}

\begin{figure}[h]
\centering
  \includegraphics[width=\textwidth]{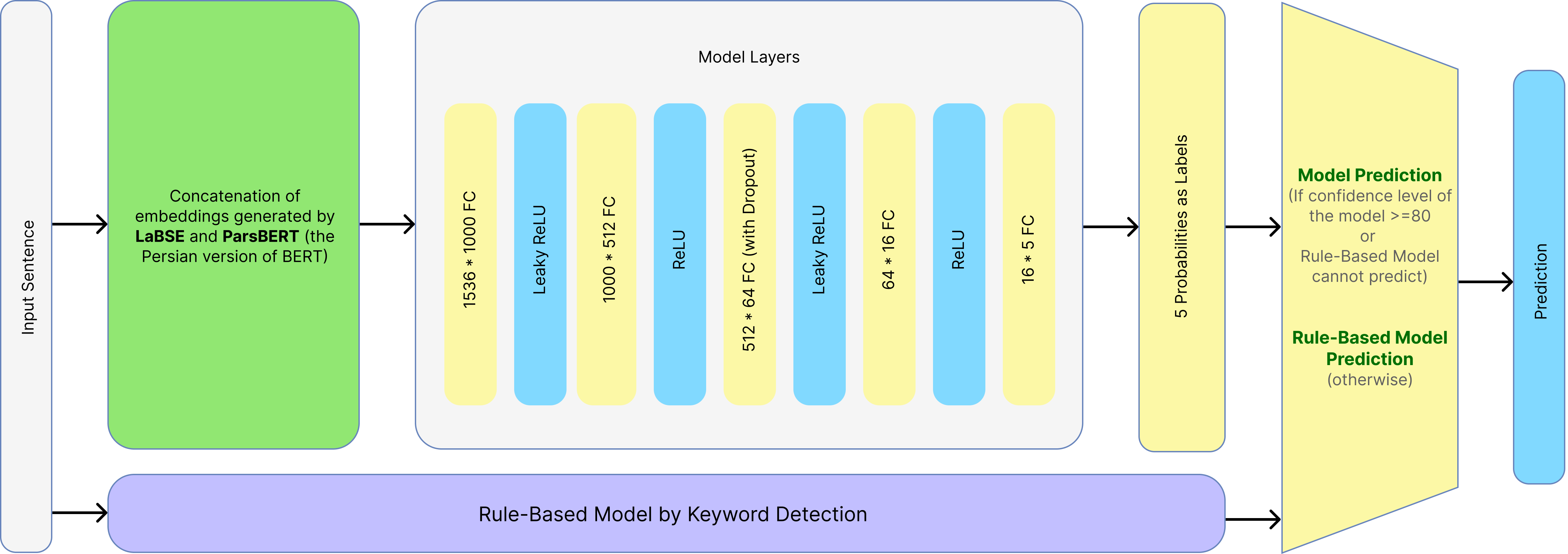}

  \caption{Architecture of the final hybrid sentiment classification model. The model takes a Persian input sentence, generates two sets of contextual embeddings using ParsBERT and LaBSE models, and concatenates the generated embeddings into a single vector. The final embedding vector is passed through a series of fully connected layers and non-linear activation functions (e.g., ReLU and Leaky ReLU) to produce a probability distribution over five sentiment classes. In parallel, a rule-based model based on keyword detection attempts to predict sentiment. The final output is selected based on a decision rule: if the neural model's confidence is $\geq 80$ or the rule-based system cannot make a prediction, the model's output is used; otherwise, the rule-based result is chosen.}

  \label{app:arch}

\end{figure}

\begin{table}[h]
\centering
\caption{Evaluation metrics of different models on the test dataset.}
\begin{tabular*}{\textwidth}{@{\extracolsep{\fill}}lcccc}
\toprule
\textbf{Classification Model} & \textbf{Accuracy} & \textbf{F1 Score} & \textbf{Precision} & \textbf{Recall} \\
\midrule
Baseline                       & 20.00\% & 20.00\% & 20.00\% & 20.00\% \\
Random Forest                  & 48.00\% & 48.90\% & 51.04\% & 47.06\% \\
Transformers Encoder-Decoder   & 51.02\% & 50.53\% & 53.12\% & 48.22\% \\
Fine-Tuned Sentiment Analyzer  & 51.49\% & 48.88\% & 49.79\% & 50.92\% \\
Our Proposed Method            & 74.08\% & 72.02\% & 72.03\% & 75.99\% \\
\bottomrule
\end{tabular*}
\label{app:accuracies}
\end{table}

\twocolumn

\onecolumn

\section{Qualitative Evaluation Results}

\begin{table*}[h!]

\centering

\caption{Qualitative metrics of the evaluation.}

\begin{tabular*}{\textwidth}{@{\extracolsep{\fill}}lcccc}

\toprule

\textbf{Theme Class} & \textbf{Overall Theme} & \textbf{Representative Sample}  \\ \midrule

\textbf{$Theme_1$} & \text{Fine User Experience} 

& \text{The user interface was elegant and very easy} \\ 

\textbf{} & \text{} 

& \text{to understand.} \\ 

\midrule

\textbf{$Theme_2$} & \text{Accurate Results} 

& \text{It was pretty accurate and close to the reality.} \\ 

\midrule

\textbf{$Theme_3$} & \text{Increasing Number of} 

& \text{I think maybe considering more tweets or images} \\ 

\textbf{} & \text{Analyzed Tweets and Images} 

& \text{can improve the system.} \\ 

\midrule

\textbf{$Theme_4$} & \text{Extending the Experiment to} 

& \text{It would be great to generalize it to various social} \\

\textbf{} & \text{Various Platforms} 

& \text{media platforms to have a more accurate understanding} \\

\textbf{} & \text{} 

& \text{of the user.} \\

\midrule

\textbf{$Theme_5$} & \text{Time-Boxed Analysis} 

& \text{I suppose that it's better to analyze tweets and images} \\

\textbf{} & \text{} 

& \text{based on their time, not a fixed number of them. For} \\

\textbf{} & \text{} 

& \text{example, they can be limited to recent 5 months.} \\

\midrule

\textbf{$Theme_6$} & \text{Comprehensive Questionnaires} 

& \text{The experiment was comprehensive with efficient} \\

\textbf{} & \text{} 

& \text{questionnaires, which did not have any extra questions.} \\

\midrule

\textbf{$Theme_7$} & \text{Transparent Procedures} 

& \text{It was pretty good that all procedures of the experiment} \\

\textbf{} & \text{} 

& \text{with details were transparent, and clearly explained in} \\

\textbf{} & \text{} 

& \text{getting my consent of participation.} \\

\midrule

\textbf{$Theme_8$} & \text{Privacy Considerations} 

& \text{I was satisfied with the privacy issues which were} \\

\textbf{} & \text{} 

& \text{considered in the procedure of the experiments.} \\

\midrule

\textbf{$Theme_9$} & \text{Self-Emotions Awareness} 

& \text{The results brought a sense of self-awareness about my} \\

\textbf{} & \text{} 

& \text{emotions in my tweets!} \\

\midrule

\textbf{$Theme_{10}$} & \text{Characteristic Type Personalization} 

& \text{I think one other thing that should be taken into account} \\

\textbf{} & \text{} 

& \text{is that consider whether the individual is introverted} \\

\textbf{} & \text{} 

& \text{or extroverted.} \\

\bottomrule

\end{tabular*}

\label{app:qualitative}

\end{table*}


\section{Statistical Test Details}

\begin{table*}[h!]

\centering

\caption{Results of the independent two-sample t-test comparing sentiment proportions between different groups.}

\begin{tabular*}{\textwidth}{@{\extracolsep{\fill}}lccccc}

\toprule

\textbf{Sentiment} & \textbf{Group 1} & \textbf{Group 2} & \textbf{T-statistic} & \textbf{P-value} & \textbf{Significance} \\ \hline

\textbf{Happy} & Images & Tweets & 13.85 & $2.43 \times 10^{-25}$ & Significant \\ \hline

\textbf{Happy} & Images & Friends & 10.87 & $7.74 \times 10^{-19}$ & Significant \\ \hline

\textbf{Happy} & Tweets & Friends & -7.16 & $1.19 \times 10^{-10}$ & Significant \\ \hline

\textbf{Sad} & Images & Tweets & -10.70 & $1.84 \times 10^{-18}$ & Significant \\ \hline

\textbf{Sad} & Images & Friends & -9.61 & $5.08 \times 10^{-16}$ & Significant \\ \hline

\textbf{Sad} & Tweets & Friends & 4.83 & $4.80 \times 10^{-06}$ & Significant \\ \hline

\textbf{Angry} & Images & Tweets & -13.61 & $7.77 \times 10^{-25}$ & Significant \\ \hline

\textbf{Angry} & Images & Friends & -14.24 & $3.55 \times 10^{-26}$ & Significant \\ \hline

\textbf{Angry} & Tweets & Friends & 2.71 & 0.0078 & Significant \\ \hline

\textbf{Neutral} & Images & Tweets & -17.64 & $4.97 \times 10^{-33}$ & Significant \\ \hline

\textbf{Neutral} & Images & Friends & -19.95 & $2.45 \times 10^{-37}$ & Significant \\ \hline

\textbf{Neutral} & Tweets & Friends & 4.00 & 0.00012 & Significant \\ \hline

\textbf{Intense Emotions} & Images & Tweets & 7.32 & $5.48 \times 10^{-11}$ & Significant \\ \hline

\textbf{Intense Emotions} & Images & Friends & 6.22 & $1.04 \times 10^{-08}$ & Significant \\ \hline

\textbf{Intense Emotions} & Tweets & Friends & -2.69 & 0.0083 & Significant \\ \bottomrule

\end{tabular*}

\label{app:statistical_tests}

\end{table*}



\label{appendix:sentiment_analysis}
{\small


\onecolumn
\setlength{\tabcolsep}{7pt}
\renewcommand{\arraystretch}{0.99}
\begin{longtable}{|c|c|c|c|c|c|c|c|c|c|c|c|c|c|c|c|c|c|c|}

\caption{Sentiment Analysis Results on Real World, Tweets, and Images of all the Participants}

\\

\hline

\textbf{A\footnote{A: Participant Number: Identifier for each individual in the study.}} & 

\textbf{B\footnote{B: Images Portion of Happy: Percentage of images categorized as 'Happy'.}} & 

\textbf{C\footnote{C: Images Portion of Sad: Percentage of images categorized as 'Sad'.}} & 

\textbf{D\footnote{D: Images Portion of Angry: Percentage of images categorized as 'Angry'.}} & 

\textbf{E\footnote{E: Images Portion of Neutral: Percentage of images categorized as 'Neutral'.}} & 

\textbf{F\footnote{F: Images Portion of Intense Emotions: Percentage of images categorized as 'Intense Emotions'.}} & 

\textbf{G\footnote{G: Tweets Portion of Happy: Percentage of tweets categorized as 'Happy'.}} & 

\textbf{H\footnote{H: Tweets Portion of Sad: Percentage of tweets categorized as 'Sad'.}} & 

\textbf{I\footnote{I: Tweets Portion of Angry: Percentage of tweets categorized as 'Angry'.}} & 

\textbf{J\footnote{J: Tweets Portion of Neutral: Percentage of tweets categorized as 'Neutral'.}} & 

\textbf{K\footnote{K: Tweets Portion of Intense Emotions: Percentage of tweets categorized as 'Intense Emotions'.}} & 

\textbf{L\footnote{L: Real World Portion of Happy: Percentage of real-world data categorized as 'Happy'.}} & 

\textbf{M\footnote{M: Real World Portion of Sad: Percentage of real-world data categorized as 'Sad'.}} & 

\textbf{N\footnote{N: Real World Portion of Angry: Percentage of real-world data categorized as 'Angry'.}} & 

\textbf{O\footnote{O: Real World Portion of Neutral: Percentage of real-world data categorized as 'Neutral'.}} & 

\textbf{P\footnote{P: Real World Portion of Intense Emotions: Percentage of real-world data categorized as 'Intense Emotions'.}} & 

\textbf{Q\footnote{Q: Distance Between Images and Real World (Earth Mover's Distance): Measure of dissimilarity between image sentiment distribution and real-world sentiment distribution.}} & 

\textbf{R\footnote{R: Distance Between Tweets and Real World (Earth Mover's Distance): Measure of dissimilarity between tweet sentiment distribution and real-world sentiment distribution.}} & 

\textbf{S\footnote{S: Distance Between Images and Tweets (Earth Mover's Distance): Measure of dissimilarity between image sentiment distribution and tweet sentiment distribution.}} \\

\hline

P1 & 0.72 & 0.03 & 0.07 & 0.0 & 0.18 & 0.12 & 0.18 & 0.14 & 0.3 & 0.26 & 0.38 & 0.11 & 0.15 & 0.1 & 0.26 & 0.3 & 0.84 & 0.15 \\ 

P2 & 0.37 & 0.03 & 0.02 & 0.13 & 0.45 & 0.2 & 0.36 & 0.18 & 0.14 & 0.12 & 0.18 & 0.17 & 0.21 & 0.22 & 0.22 & 0.24 & 0.72 & 0.48 \\ 

P3 & 0.71 & 0.07 & 0.0 & 0.0 & 0.22 & 0.04 & 0.24 & 0.24 & 0.31 & 0.16 & 0.17 & 0.23 & 0.2 & 0.23 & 0.16 & 0.04 & 0.73 & 0.2 \\ 

P4 & 0.95 & 0.05 & 0.0 & 0.0 & 0.0 & 0.14 & 0.3 & 0.14 & 0.26 & 0.16 & 0.17 & 0.25 & 0.16 & 0.24 & 0.18 & -0.39 & 0.86 & -0.3 \\ 

P5 & 0.78 & 0.0 & 0.01 & 0.0 & 0.21 & 0.18 & 0.16 & 0.24 & 0.22 & 0.2 & 0.28 & 0.15 & 0.17 & 0.24 & 0.17 & 0.0 & 0.89 & -0.08 \\ 

P6 & 0.25 & 0.27 & 0.1 & 0.22 & 0.16 & 0.1 & 0.08 & 0.2 & 0.28 & 0.34 & 0.32 & 0.12 & 0.12 & 0.17 & 0.28 & 0.81 & 0.88 & 0.8 \\ 

P7 & 0.63 & 0.0 & 0.0 & 0.0 & 0.37 & 0.24 & 0.24 & 0.08 & 0.28 & 0.16 & 0.29 & 0.18 & 0.14 & 0.16 & 0.23 & 0.03 & 0.86 & 0.04 \\ 

P8 & 0.82 & 0.03 & 0.01 & 0.03 & 0.11 & 0.16 & 0.2 & 0.18 & 0.22 & 0.24 & 0.29 & 0.12 & 0.13 & 0.18 & 0.28 & -0.06 & 0.78 & -0.15 \\ 

P9 & 0.5 & 0.08 & 0.05 & 0.07 & 0.3 & 0.14 & 0.38 & 0.14 & 0.1 & 0.24 & 0.19 & 0.25 & 0.12 & 0.22 & 0.22 & 0.34 & 0.7 & 0.64 \\ 

P10 & 0.26 & 0.01 & 0.01 & 0.0 & 0.72 & 0.04 & 0.26 & 0.38 & 0.18 & 0.14 & 0.16 & 0.29 & 0.19 & 0.16 & 0.2 & 0.02 & 0.7 & 0.31 \\ 

P11 & 0.5 & 0.09 & 0.02 & 0.08 & 0.31 & 0.28 & 0.36 & 0.12 & 0.14 & 0.1 & 0.2 & 0.2 & 0.1 & 0.28 & 0.22 & 0.37 & 0.72 & 0.65 \\ 

P12 & 0.28 & 0.03 & 0.01 & 0.0 & 0.68 & 0.16 & 0.24 & 0.1 & 0.26 & 0.24 & 0.3 & 0.2 & 0.06 & 0.2 & 0.23 & 0.15 & 0.82 & 0.07 \\ 

P13 & 0.36 & 0.13 & 0.13 & 0.0 & 0.38 & 0.09 & 0.09 & 0.29 & 0.4 & 0.14 & 0.21 & 0.18 & 0.16 & 0.25 & 0.2 & 0.43 & 0.55 & 0.77 \\ 

P14 & 0.58 & 0.13 & 0.03 & 0.0 & 0.26 & 0.16 & 0.16 & 0.14 & 0.4 & 0.14 & 0.23 & 0.13 & 0.2 & 0.25 & 0.2 & 0.27 & 0.67 & 0.44 \\ 

P15 & 0.76 & 0.01 & 0.02 & 0.01 & 0.2 & 0.2 & 0.2 & 0.22 & 0.24 & 0.14 & 0.21 & 0.18 & 0.14 & 0.22 & 0.24 & -0.03 & 0.96 & -0.04 \\ 

P16 & 0.22 & 0.15 & 0.12 & 0.01 & 0.5 & 0.14 & 0.12 & 0.3 & 0.34 & 0.1 & 0.2 & 0.15 & 0.13 & 0.33 & 0.19 & 0.62 & 0.78 & 0.66 \\ 

P17 & 0.86 & 0.03 & 0.0 & 0.0 & 0.11 & 0.16 & 0.22 & 0.1 & 0.28 & 0.24 & 0.12 & 0.31 & 0.18 & 0.18 & 0.2 & -0.09 & 0.84 & -0.15 \\ 

P18 & 0.69 & 0.03 & 0.01 & 0.0 & 0.27 & 0.08 & 0.16 & 0.12 & 0.4 & 0.24 & 0.29 & 0.11 & 0.09 & 0.23 & 0.28 & 0.2 & 0.75 & 0.36 \\ 

P19 & 0.22 & 0.26 & 0.12 & 0.1 & 0.31 & 0.14 & 0.08 & 0.28 & 0.3 & 0.2 & 0.15 & 0.22 & 0.24 & 0.18 & 0.21 & 0.77 & 0.76 & 0.91 \\ 

P20 & 0.46 & 0.07 & 0.06 & 0.01 & 0.4 & 0.1 & 0.26 & 0.26 & 0.16 & 0.22 & 0.22 & 0.21 & 0.16 & 0.16 & 0.24 & 0.2 & 0.87 & 0.31 \\ 

P21 & 0.56 & 0.03 & 0.04 & 0.01 & 0.36 & 0.2 & 0.2 & 0.2 & 0.14 & 0.26 & 0.25 & 0.13 & 0.13 & 0.25 & 0.25 & 0.15 & 0.8 & 0.07 \\ 

P22 & 0.64 & 0.03 & 0.01 & 0.0 & 0.32 & 0.23 & 0.2 & 0.14 & 0.11 & 0.31 & 0.22 & 0.19 & 0.15 & 0.18 & 0.27 & 0.05 & 0.85 & 0.17 \\ 

P23 & 0.64 & 0.0 & 0.0 & 0.0 & 0.36 & 0.14 & 0.26 & 0.18 & 0.2 & 0.22 & 0.22 & 0.16 & 0.1 & 0.25 & 0.26 & 0.03 & 0.88 & -0.04 \\ 

P24 & 0.45 & 0.06 & 0.05 & 0.01 & 0.42 & 0.06 & 0.22 & 0.32 & 0.12 & 0.28 & 0.28 & 0.16 & 0.18 & 0.18 & 0.21 & 0.21 & 0.68 & 0.44 \\ 

P25 & 0.44 & 0.07 & 0.05 & 0.1 & 0.34 & 0.17 & 0.17 & 0.08 & 0.25 & 0.33 & 0.25 & 0.17 & 0.15 & 0.18 & 0.25 & 0.43 & 0.84 & 0.6 \\ 

P26 & 0.18 & 0.01 & 0.01 & 0.0 & 0.8 & 0.23 & 0.23 & 0.14 & 0.34 & 0.06 & 0.3 & 0.2 & 0.16 & 0.12 & 0.22 & 0.0 & 0.84 & 0.07 \\ 

P27 & 0.94 & 0.0 & 0.0 & 0.0 & 0.06 & 0.15 & 0.32 & 0.09 & 0.35 & 0.09 & 0.29 & 0.14 & 0.19 & 0.19 & 0.2 & -0.3 & 0.63 & -0.17 \\ 

P28 & 0.39 & 0.05 & 0.06 & 0.03 & 0.48 & 0.28 & 0.16 & 0.24 & 0.2 & 0.12 & 0.0 & 0.0 & 0.0 & 0.0 & 0.0 & 0.0 & 0.0 & 0.31 \\ 

P29 & 0.8 & 0.09 & 0.03 & 0.0 & 0.08 & 0.16 & 0.22 & 0.24 & 0.16 & 0.22 & 0.23 & 0.14 & 0.23 & 0.11 & 0.28 & -0.03 & 0.86 & -0.12 \\ 

P30 & 0.51 & 0.08 & 0.08 & 0.02 & 0.31 & 0.28 & 0.18 & 0.18 & 0.12 & 0.24 & 0.27 & 0.16 & 0.1 & 0.19 & 0.29 & 0.46 & 0.91 & 0.4 \\ 

P31 & 0.38 & 0.45 & 0.0 & 0.0 & 0.18 & 0.28 & 0.3 & 0.08 & 0.18 & 0.16 & 0.23 & 0.24 & 0.14 & 0.09 & 0.3 & 0.43 & 0.88 & 0.51 \\ 

P32 & 0.46 & 0.09 & 0.05 & 0.04 & 0.36 & 0.17 & 0.23 & 0.15 & 0.25 & 0.21 & 0.25 & 0.18 & 0.13 & 0.21 & 0.23 & 0.32 & 0.96 & 0.31 \\ 

P33 & 0.36 & 0.1 & 0.13 & 0.12 & 0.29 & 0.24 & 0.26 & 0.1 & 0.26 & 0.14 & 0.32 & 0.12 & 0.16 & 0.13 & 0.27 & 0.88 & 0.82 & 0.74 \\ 

P34 & 0.37 & 0.17 & 0.09 & 0.11 & 0.26 & 0.08 & 0.2 & 0.18 & 0.16 & 0.38 & 0.23 & 0.29 & 0.14 & 0.19 & 0.16 & 0.76 & 0.81 & 0.86 \\ 

P35 & 0.57 & 0.01 & 0.01 & 0.0 & 0.41 & 0.14 & 0.16 & 0.16 & 0.42 & 0.12 & 0.18 & 0.17 & 0.22 & 0.24 & 0.19 & -0.03 & 0.64 & 0.2 \\ 

P36 & 0.32 & 0.08 & 0.01 & 0.0 & 0.59 & 0.14 & 0.26 & 0.12 & 0.28 & 0.2 & 0.15 & 0.32 & 0.09 & 0.2 & 0.23 & 0.29 & 0.88 & 0.26 \\ 

P37 & 0.23 & 0.01 & 0.01 & 0.03 & 0.72 & 0.22 & 0.12 & 0.2 & 0.36 & 0.1 & 0.26 & 0.17 & 0.17 & 0.24 & 0.17 & 0.07 & 0.73 & 0.26 \\ 

P38 & 0.46 & 0.05 & 0.08 & 0.08 & 0.33 & 0.08 & 0.3 & 0.04 & 0.1 & 0.48 & 0.15 & 0.27 & 0.12 & 0.2 & 0.25 & 0.47 & 0.49 & 0.92 \\ 

P39 & 0.17 & 0.05 & 0.07 & 0.17 & 0.54 & 0.22 & 0.24 & 0.18 & 0.24 & 0.12 & 0.23 & 0.15 & 0.22 & 0.23 & 0.17 & 0.38 & 0.94 & 0.39 \\ 

P40 & 0.36 & 0.12 & 0.04 & 0.02 & 0.47 & 0.18 & 0.2 & 0.16 & 0.18 & 0.28 & 0.24 & 0.17 & 0.11 & 0.28 & 0.21 & 0.38 & 0.86 & 0.31 \\ 

P41 & 0.42 & 0.19 & 0.02 & 0.0 & 0.36 & 0.24 & 0.24 & 0.2 & 0.22 & 0.1 & 0.39 & 0.1 & 0.12 & 0.12 & 0.27 & 0.6 & 0.63 & 0.38 \\ 

P42 & 0.83 & 0.03 & 0.0 & 0.0 & 0.14 & 0.08 & 0.08 & 0.18 & 0.5 & 0.16 & 0.36 & 0.11 & 0.1 & 0.1 & 0.34 & 0.05 & 0.61 & 0.34 \\ 

P43 & 0.48 & 0.01 & 0.02 & 0.04 & 0.46 & 0.1 & 0.36 & 0.12 & 0.16 & 0.26 & 0.19 & 0.24 & 0.15 & 0.26 & 0.17 & 0.12 & 0.75 & 0.37 \\ 

P44 & 0.36 & 0.05 & 0.07 & 0.05 & 0.47 & 0.42 & 0.1 & 0.04 & 0.28 & 0.16 & 0.35 & 0.08 & 0.06 & 0.37 & 0.14 & 0.78 & 0.82 & 0.72 \\ 

P45 & 0.66 & 0.01 & 0.0 & 0.0 & 0.33 & 0.06 & 0.18 & 0.06 & 0.5 & 0.2 & 0.29 & 0.12 & 0.07 & 0.31 & 0.2 & 0.23 & 0.62 & 0.41 \\ 

P46 & 0.44 & 0.25 & 0.01 & 0.0 & 0.31 & 0.16 & 0.26 & 0.08 & 0.34 & 0.16 & 0.27 & 0.08 & 0.14 & 0.17 & 0.34 & 0.57 & 0.96 & 0.53 \\ 

P47 & 0.9 & 0.03 & 0.01 & 0.0 & 0.06 & 0.12 & 0.3 & 0.2 & 0.28 & 0.1 & 0.13 & 0.31 & 0.2 & 0.2 & 0.17 & -0.18 & 0.83 & -0.2 \\ 

P48 & 0.31 & 0.01 & 0.01 & 0.0 & 0.67 & 0.1 & 0.36 & 0.1 & 0.22 & 0.22 & 0.23 & 0.21 & 0.1 & 0.33 & 0.13 & 0.16 & 0.92 & 0.2 \\ 

P49 & 0.36 & 0.2 & 0.06 & 0.1 & 0.27 & 0.18 & 0.08 & 0.08 & 0.36 & 0.3 & 0.32 & 0.13 & 0.11 & 0.15 & 0.3 & 0.79 & 0.84 & 0.9 \\ 

P50 & 0.64 & 0.06 & 0.08 & 0.11 & 0.11 & 0.16 & 0.34 & 0.16 & 0.3 & 0.04 & 0.26 & 0.17 & 0.16 & 0.14 & 0.27 & 0.26 & 0.78 & 0.36 \\ 

P51 & 0.34 & 0.19 & 0.08 & 0.0 & 0.4 & 0.18 & 0.2 & 0.24 & 0.24 & 0.14 & 0.23 & 0.18 & 0.11 & 0.25 & 0.23 & 0.49 & 0.92 & 0.49 \\ 

P52 & 0.53 & 0.03 & 0.06 & 0.02 & 0.36 & 0.26 & 0.1 & 0.08 & 0.34 & 0.22 & 0.31 & 0.13 & 0.07 & 0.28 & 0.22 & 0.38 & 0.9 & 0.42 \\ 

P53 & 0.52 & 0.04 & 0.08 & 0.1 & 0.26 & 0.24 & 0.08 & 0.16 & 0.38 & 0.14 & 0.21 & 0.15 & 0.21 & 0.26 & 0.18 & 0.37 & 0.69 & 0.68 \\ 

P54 & 0.39 & 0.19 & 0.02 & 0.06 & 0.34 & 0.1 & 0.08 & 0.1 & 0.42 & 0.3 & 0.31 & 0.15 & 0.14 & 0.17 & 0.23 & 0.57 & 0.64 & 0.74 \\ 

P55 & 0.58 & 0.06 & 0.03 & 0.02 & 0.31 & 0.24 & 0.12 & 0.18 & 0.22 & 0.24 & 0.37 & 0.1 & 0.09 & 0.15 & 0.28 & 0.53 & 0.64 & 0.17 \\ 

P56 & 0.33 & 0.15 & 0.08 & 0.03 & 0.4 & 0.24 & 0.14 & 0.14 & 0.14 & 0.34 & 0.14 & 0.25 & 0.17 & 0.19 & 0.25 & 0.52 & 0.82 & 0.66 \\ 

P57 & 0.46 & 0.04 & 0.02 & 0.0 & 0.48 & 0.14 & 0.26 & 0.26 & 0.21 & 0.12 & 0.27 & 0.17 & 0.13 & 0.16 & 0.28 & 0.2 & 0.91 & 0.17 \\ 

P58 & 0.05 & 0.17 & 0.02 & 0.06 & 0.7 & 0.14 & 0.28 & 0.16 & 0.14 & 0.28 & 0.24 & 0.24 & 0.05 & 0.21 & 0.25 & 0.1 & 0.69 & 0.16 \\ 

P59 & 0.74 & 0.04 & 0.01 & 0.01 & 0.2 & 0.14 & 0.34 & 0.08 & 0.06 & 0.38 & 0.31 & 0.12 & 0.1 & 0.16 & 0.31 & 0.14 & 0.8 & 0.28 \\ 

P60 & 0.41 & 0.12 & 0.09 & 0.05 & 0.33 & 0.24 & 0.28 & 0.18 & 0.14 & 0.16 & 0.22 & 0.24 & 0.15 & 0.17 & 0.22 & 0.43 & 0.88 & 0.56 \\ 

P61 & 0.23 & 0.01 & 0.01 & 0.0 & 0.75 & 0.2 & 0.26 & 0.22 & 0.14 & 0.18 & 0.16 & 0.16 & 0.19 & 0.2 & 0.28 & 0.0 & 0.91 & 0.0 \\ 

P62 & 0.08 & 0.24 & 0.19 & 0.3 & 0.19 & 0.04 & 0.29 & 0.36 & 0.18 & 0.14 & 0.21 & 0.21 & 0.18 & 0.21 & 0.19 & 0.76 & 0.55 & 0.79 \\ 

P63 & 0.43 & 0.19 & 0.16 & 0.07 & 0.15 & 0.1 & 0.34 & 0.08 & 0.24 & 0.24 & 0.2 & 0.17 & 0.19 & 0.19 & 0.26 & 0.65 & 0.64 & 0.72 \\ 

P64 & 1.0 & 0.0 & 0.0 & 0.0 & 0.0 & 0.14 & 0.09 & 0.06 & 0.49 & 0.23 & 0.28 & 0.15 & 0.07 & 0.25 & 0.26 & -0.44 & 0.58 & -0.02 \\ 

P65 & 0.71 & 0.07 & 0.02 & 0.02 & 0.18 & 0.16 & 0.2 & 0.12 & 0.32 & 0.2 & 0.29 & 0.09 & 0.11 & 0.17 & 0.34 & 0.26 & 0.78 & 0.21 \\ 

P66 & 0.37 & 0.03 & 0.03 & 0.01 & 0.56 & 0.25 & 0.2 & 0.05 & 0.15 & 0.35 & 0.24 & 0.18 & 0.06 & 0.28 & 0.24 & 0.17 & 0.84 & 0.33 \\ 

P67 & 0.99 & 0.01 & 0.0 & 0.0 & 0.0 & 0.32 & 0.16 & 0.18 & 0.26 & 0.08 & 0.29 & 0.28 & 0.1 & 0.12 & 0.21 & -0.39 & 0.85 & -0.34 \\ 

P68 & 0.15 & 0.06 & 0.04 & 0.08 & 0.67 & 0.1 & 0.06 & 0.28 & 0.3 & 0.26 & 0.28 & 0.16 & 0.07 & 0.24 & 0.25 & 0.21 & 0.86 & 0.25 \\ 

P69 & 0.28 & 0.03 & 0.01 & 0.0 & 0.68 & 0.08 & 0.24 & 0.14 & 0.3 & 0.24 & 0.34 & 0.17 & 0.05 & 0.18 & 0.25 & 0.27 & 0.82 & 0.15 \\ 

P70 & 0.56 & 0.31 & 0.02 & 0.02 & 0.09 & 0.4 & 0.12 & 0.16 & 0.2 & 0.12 & 0.27 & 0.16 & 0.14 & 0.15 & 0.29 & 0.36 & 0.77 & 0.46 \\ 

P71 & 0.34 & 0.2 & 0.13 & 0.09 & 0.25 & 0.06 & 0.12 & 0.06 & 0.51 & 0.25 & 0.26 & 0.23 & 0.1 & 0.17 & 0.23 & 0.82 & 0.46 & 0.64 \\ 

P72 & 0.13 & 0.02 & 0.05 & 0.03 & 0.77 & 0.12 & 0.32 & 0.14 & 0.22 & 0.2 & 0.21 & 0.34 & 0.08 & 0.08 & 0.29 & 0.13 & 0.8 & 0.09 \\ 

P73 & 0.32 & 0.12 & 0.08 & 0.1 & 0.38 & 0.26 & 0.18 & 0.22 & 0.24 & 0.1 & 0.25 & 0.13 & 0.13 & 0.22 & 0.27 & 0.64 & 0.9 & 0.6 \\ 

P74 & 0.58 & 0.01 & 0.02 & 0.0 & 0.39 & 0.05 & 0.41 & 0.05 & 0.14 & 0.36 & 0.16 & 0.2 & 0.17 & 0.17 & 0.3 & 0.06 & 0.47 & 0.58 \\ 

P75 & 0.34 & 0.19 & 0.05 & 0.1 & 0.33 & 0.22 & 0.22 & 0.1 & 0.12 & 0.34 & 0.19 & 0.24 & 0.14 & 0.18 & 0.25 & 0.65 & 0.76 & 0.78 \\ 

P76 & 0.54 & 0.06 & 0.04 & 0.02 & 0.34 & 0.06 & 0.52 & 0.2 & 0.04 & 0.18 & 0.24 & 0.23 & 0.14 & 0.12 & 0.26 & 0.25 & 0.48 & 0.68 \\ 

P77 & 0.17 & 0.08 & 0.17 & 0.15 & 0.43 & 0.2 & 0.3 & 0.1 & 0.24 & 0.16 & 0.23 & 0.19 & 0.17 & 0.24 & 0.17 & 0.62 & 0.84 & 0.74 \\ 

P78 & 0.08 & 0.12 & 0.11 & 0.01 & 0.68 & 0.5 & 0.1 & 0.08 & 0.2 & 0.12 & 0.33 & 0.1 & 0.07 & 0.19 & 0.31 & 0.3 & 0.64 & 0.63 \\ 

P79 & 0.48 & 0.1 & 0.02 & 0.02 & 0.38 & 0.1 & 0.32 & 0.12 & 0.32 & 0.14 & 0.17 & 0.27 & 0.15 & 0.25 & 0.17 & 0.3 & 0.74 & 0.56 \\ 

P80 & 0.42 & 0.14 & 0.06 & 0.04 & 0.33 & 0.14 & 0.16 & 0.22 & 0.28 & 0.2 & 0.29 & 0.19 & 0.11 & 0.16 & 0.25 & 0.56 & 0.92 & 0.48 \\ 

P81 & 0.94 & 0.0 & 0.0 & 0.0 & 0.06 & 0.1 & 0.3 & 0.1 & 0.24 & 0.26 & 0.24 & 0.19 & 0.17 & 0.22 & 0.17 & -0.39 & 0.71 & -0.28 \\ 

P82 & 0.47 & 0.1 & 0.14 & 0.02 & 0.28 & 0.1 & 0.4 & 0.16 & 0.12 & 0.22 & 0.21 & 0.17 & 0.14 & 0.26 & 0.22 & 0.47 & 0.72 & 0.75 \\ 

P83 & 0.43 & 0.03 & 0.02 & 0.01 & 0.51 & 0.26 & 0.24 & 0.1 & 0.14 & 0.26 & 0.28 & 0.17 & 0.14 & 0.18 & 0.23 & 0.14 & 0.82 & 0.15 \\ 

P84 & 0.49 & 0.05 & 0.04 & 0.0 & 0.42 & 0.1 & 0.34 & 0.14 & 0.28 & 0.14 & 0.18 & 0.32 & 0.23 & 0.14 & 0.14 & 0.26 & 0.84 & 0.42 \\ 

P85 & 0.37 & 0.06 & 0.01 & 0.01 & 0.55 & 0.06 & 0.44 & 0.1 & 0.16 & 0.24 & 0.25 & 0.17 & 0.17 & 0.18 & 0.22 & 0.1 & 0.58 & 0.52 \\ 

P86 & 0.57 & 0.07 & 0.02 & 0.01 & 0.33 & 0.14 & 0.1 & 0.12 & 0.32 & 0.32 & 0.29 & 0.1 & 0.05 & 0.29 & 0.26 & 0.37 & 0.75 & 0.48 \\ 

P87 & 0.52 & 0.09 & 0.04 & 0.03 & 0.32 & 0.08 & 0.48 & 0.14 & 0.18 & 0.12 & 0.23 & 0.15 & 0.18 & 0.2 & 0.25 & 0.27 & 0.53 & 0.63 \\ 

P88 & 0.43 & 0.07 & 0.03 & 0.15 & 0.32 & 0.19 & 0.31 & 0.08 & 0.31 & 0.11 & 0.15 & 0.27 & 0.15 & 0.2 & 0.22 & 0.48 & 0.74 & 0.74 \\ 

P89 & 0.59 & 0.06 & 0.07 & 0.04 & 0.24 & 0.14 & 0.22 & 0.18 & 0.22 & 0.24 & 0.33 & 0.11 & 0.13 & 0.2 & 0.24 & 0.46 & 0.79 & 0.26 \\ 

P90 & 0.9 & 0.01 & 0.0 & 0.0 & 0.09 & 0.52 & 0.12 & 0.16 & 0.12 & 0.08 & 0.21 & 0.21 & 0.18 & 0.26 & 0.14 & -0.28 & 0.48 & 0.24 \\ 

P91 & 0.44 & 0.11 & 0.08 & 0.06 & 0.31 & 0.1 & 0.1 & 0.44 & 0.18 & 0.18 & 0.26 & 0.16 & 0.1 & 0.29 & 0.19 & 0.6 & 0.7 & 0.74 \\ 

P92 & 0.79 & 0.01 & 0.0 & 0.0 & 0.2 & 0.2 & 0.44 & 0.08 & 0.18 & 0.1 & 0.22 & 0.22 & 0.12 & 0.22 & 0.21 & -0.13 & 0.56 & 0.29 \\ 

P93 & 0.45 & 0.1 & 0.11 & 0.03 & 0.3 & 0.14 & 0.3 & 0.18 & 0.2 & 0.18 & 0.19 & 0.24 & 0.11 & 0.29 & 0.16 & 0.55 & 0.89 & 0.48 \\ 

P94 & 0.59 & 0.05 & 0.03 & 0.02 & 0.31 & 0.18 & 0.32 & 0.16 & 0.18 & 0.16 & 0.26 & 0.22 & 0.13 & 0.26 & 0.13 & 0.24 & 0.76 & 0.19 \\ 

P95 & 0.46 & 0.18 & 0.05 & 0.08 & 0.24 & 0.14 & 0.2 & 0.28 & 0.2 & 0.18 & 0.1 & 0.32 & 0.27 & 0.18 & 0.13 & 0.72 & 0.78 & 0.57 \\ 

P96 & 0.38 & 0.06 & 0.1 & 0.01 & 0.45 & 0.08 & 0.12 & 0.08 & 0.34 & 0.38 & 0.17 & 0.2 & 0.11 & 0.3 & 0.22 & 0.38 & 0.6 & 0.78 \\ 

P97 & 0.19 & 0.01 & 0.02 & 0.0 & 0.78 & 0.16 & 0.3 & 0.14 & 0.22 & 0.18 & 0.3 & 0.17 & 0.15 & 0.17 & 0.21 & 0.03 & 0.96 & 0.03 \\ 

P98 & 0.18 & 0.42 & 0.04 & 0.19 & 0.17 & 0.2 & 0.28 & 0.04 & 0.2 & 0.28 & 0.26 & 0.17 & 0.07 & 0.24 & 0.26 & 0.68 & 0.86 & 0.72 \\ 

P99 & 0.65 & 0.06 & 0.03 & 0.01 & 0.25 & 0.12 & 0.22 & 0.1 & 0.3 & 0.26 & 0.25 & 0.12 & 0.16 & 0.26 & 0.22 & 0.2 & 0.88 & 0.29 \\ 

P100 & 0.34 & 0.1 & 0.06 & 0.03 & 0.48 & 0.28 & 0.22 & 0.04 & 0.22 & 0.24 & 0.47 & 0.12 & 0.09 & 0.09 & 0.22 & 0.77 & 0.5 & 0.41 \\ 

P101 & 0.42 & 0.05 & 0.17 & 0.02 & 0.33 & 0.16 & 0.22 & 0.1 & 0.34 & 0.18 & 0.35 & 0.15 & 0.08 & 0.27 & 0.15 & 0.68 & 0.88 & 0.6 \\ 

P102 & 0.73 & 0.02 & 0.02 & 0.0 & 0.23 & 0.14 & 0.1 & 0.26 & 0.36 & 0.14 & 0.21 & 0.15 & 0.21 & 0.21 & 0.22 & -0.06 & 0.62 & 0.26 \\ 

P103 & 0.47 & 0.04 & 0.03 & 0.09 & 0.37 & 0.14 & 0.3 & 0.14 & 0.12 & 0.3 & 0.26 & 0.18 & 0.11 & 0.13 & 0.32 & 0.48 & 0.88 & 0.52 \\ 

P104 & 0.48 & 0.1 & 0.03 & 0.22 & 0.17 & 0.12 & 0.16 & 0.32 & 0.16 & 0.24 & 0.27 & 0.14 & 0.14 & 0.19 & 0.26 & 0.57 & 0.86 & 0.66 \\ 

P105 & 0.44 & 0.02 & 0.0 & 0.0 & 0.54 & 0.22 & 0.04 & 0.14 & 0.32 & 0.28 & 0.25 & 0.18 & 0.17 & 0.14 & 0.26 & 0.05 & 0.74 & 0.24 \\

\hline


\end{longtable}



\vfill

\end{document}